# Polarization-entangled photon pair sources based on spontaneous four wave mixing assisted by polarization mode dispersion

Pisek Kultavewuti[1,*], Eric Y. Zhu[1], Xingxing Xing[1], Li Qian[1], Vincenzo Pusino[2], Marc Sorel[2], J. Stewart Aitchison[1]

[1]Department of Electrical and Computer Engineering, University of Toronto, 10 King's College Road, Toronto, Ontario M5S 3G4, Canada
[2]School of Engineering, University of Glasgow, Glasgow G12 8QQ, Scotland, UK
Correspondence and requests for materials should be addressed to P.K. (email: pisek.kultavewuti@mail.utoronto.ca) or J.S.A (email: stewart.aitchison@utoronto.ca)

Photonic-based qubits and integrated photonic circuits have enabled demonstrations of quantum information processing (QIP) that promises to transform the way in which we compute and communicate. To that end, sources of polarization-entangled photon pair states are an important enabling technology, especially for polarization-based protocols. However, such states are difficult to prepare in an integrated photonic circuit. Scalable semiconductor sources typically rely on nonlinear optical effects where polarization mode dispersion (PMD) degrades entanglement. Here, we directly generate polarization-entangled states in an AlGaAs waveguide, aided by the PMD and without any compensation steps. We perform quantum state tomography and report a raw concurrence as high as 0.91±0.01 observed in the 1100-nm-wide waveguide. The scheme allows direct Bell state generation with an observed maximum fidelity of 0.90±0.01 from the 800-nm-wide waveguide. Our demonstration paves the way for sources that allow for the implementation of polarization-encoded protocols in large-scale quantum photonic circuits.

## Introduction

Integrated sources of polarization-entangled photons are of paramount importance for the utilization of quantum information science in real world applications. Polarization-entangled photons have been used to realize quantum teleportation[1,2], implement quantum logic and computation operations[3,4], and to simulate quantum walk[5,6], to name a few. However, in most of these demonstrations, nonlinear crystals (e.g. BBO, periodically poled LiNbO$_3$, or KTP) were used in the preparation and manipulation of the polarization-entangled states, and thereby severely limits scalability of the quantum information processing. As a result there is an urgent need for a scalable integrated platform which can efficiently produce and manipulate polarization-entangled photons[7–9]. Integrated semiconductor waveguides are a promising platform as they offer the prospect of integrating multiple functionalities on to the same chip as well as the inherent interferometric stability associated with integration. Unfortunately, producing polarization-entangled states in an integrated semiconductor waveguide is still challenging predominantly due to the polarization mode dispersion (PMD) that degrades the degree of entanglement[10,11].

Most polarization-entanglement sources that are based on semiconductor waveguides rely on either the second-order $\chi^{(2)}$ process or the third-order $\chi^{(3)}$ process, in which pump photons (one photon for $\chi^{(2)}$ and two for $\chi^{(3)}$) spontaneously decay into two correlated photons. The former process is commonly known as spontaneous parametric down-conversion (SPDC) and typically exploited in III-V materials such as AlGaAs[12–16]. However, using the $\chi^{(2)}$ process requires phase matching which is challenging in materials with no natural birefringence. A sophisticated waveguide called a Bragg-reflection waveguide (BRW)[12–15] or a quasi-phase matched waveguide (QPMW)[16] can mitigate the phase matching requirement but at a price of increased complexity of the waveguide structure. On the other hand, the $\chi^{(3)}$ process, specifically spontaneous four-wave mixing (SFWM), can meet the phase matching requirement more readily and therefore allows a simpler waveguide design. Crystalline silicon waveguide devices have been used to generate polarization-entangled photon pairs via the SFWM process[17–20]. Only one of the two polarizations of the fundamental waveguide modes is exploited, and it is therefore mandatory to devise a polarization rotation step either through an additional fiber polarization-controlled Sagnac loop[17,20] or a micro-scaled polarization rotator[19]. Lv et al[21] showed the possibility of utilizing both modes in parallel SFWM processes to generate polarization-entangled states in silicon waveguides. However, the PMD is still a major limitation and the demonstrated device is far from optimal.

In this work, we demonstrate a simple way of generating polarization-entangled photon pairs via orthogonally-polarized SFWM processes and show that an optimal, non-vanishing polarization mode dispersion is a necessity to generate such highly entangled states. The technique does not require any compensation steps, either on-chip or off-chip. The source is based on a deeply etched AlGaAs waveguide which is a promising platform for integrated quantum photonics. Proven by quantum state tomography measurements of different waveguides on the same chip, the source not only exhibits a high raw concurrence of 0.91±0.01 but also is capable of directly producing Bell states with a maximum fidelity of 0.90±0.01.

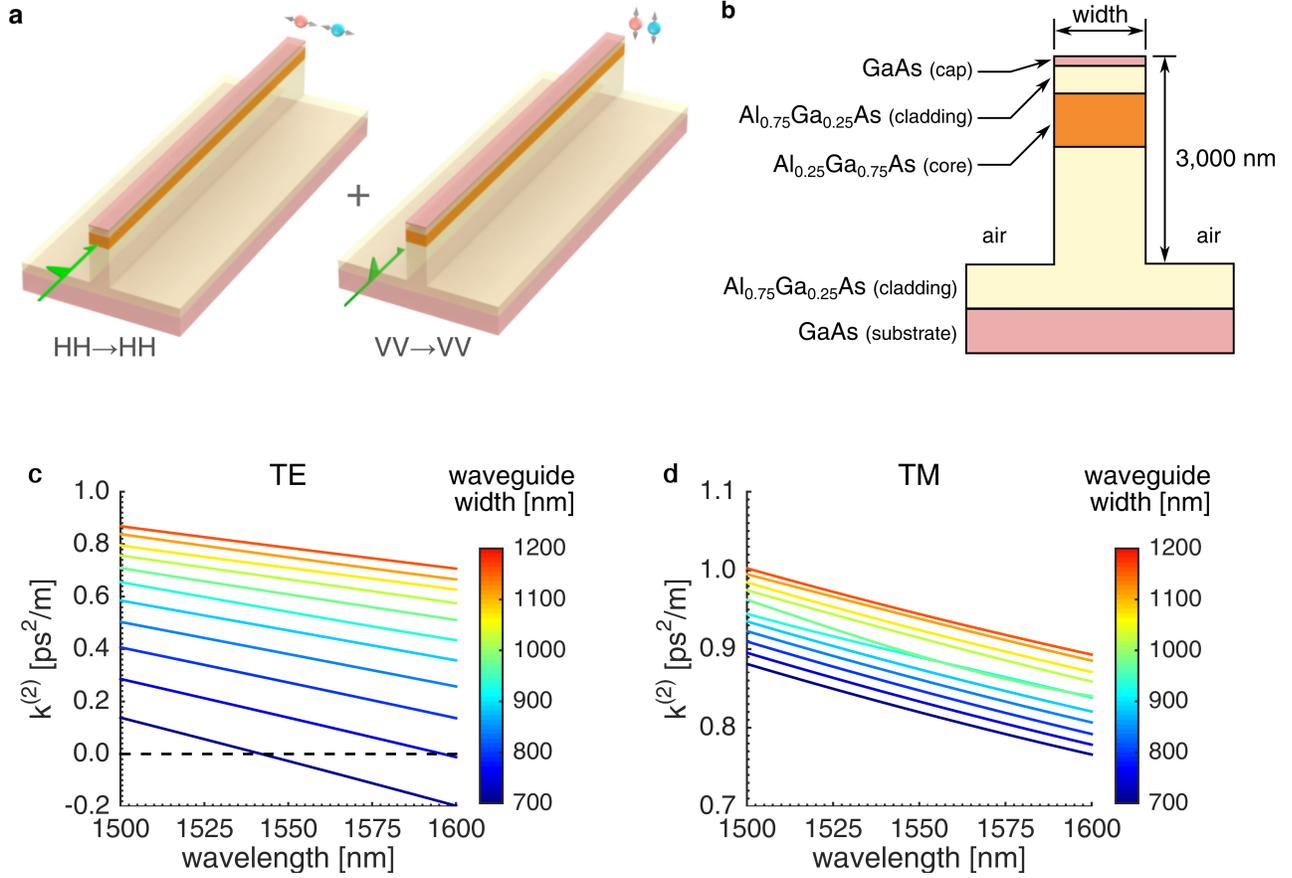

Figure 1. (a) The two orthogonally pumped SFWM processes that drive the entanglement generation. (b) A cross section of a deeply etched AlGaAs polarization-entangled photon pair source, detailing the materials composing the structure. (c) and (d) The dispersion coefficients $k^{(2)}$ of the TE and TM waveguide modes. The zero-dispersion wavelength near 1,550 nm is achieved for the TE mode of the waveguide width of 700 nm.

## Results

**Basic concepts and device design**. Our entanglement scheme relies on two independent SFWM processes. The entanglement originates from the lack of knowledge of which SFWM processes create a photon pair. We implement the scheme in a deeply etched AlGaAs waveguide that supports two fundamental modes at the telecommunications wavelengths, namely: the Transverse Electric (TE) and the Transverse Magnetic (TM) modes. The TE (TM) mode has its major electric field component orientated horizontally (vertically) with respect to the growth direction of the wafer and is interchangeably represented by a letter $H$ ($V$) for its polarization state. In such a waveguide, SFWM processes with co-polarized interacting photons are preferentially phase matched in the spectral region of near degeneracy. Therefore, it is more likely that both the signal and the idler photons are generated in the $H$ state from two pump photons which are also in the $H$ state, and similarly for the process involving photons all in the *V* state. Essentially, we obtain at the output of the waveguide a polarization-entangled state (see Fig. 1a) which can be described

$$|\psi\rangle = a|HH\rangle + be^{i\theta}|VV\rangle \tag{1}$$

where the first and second letters in the ket signifies the polarization states of the signal and idler photons respectively, the coefficients *a* and *b* are real numbers, and *θ* is the relative phase difference between the |*HH*⟩ and |*VV*⟩.

We fabricated waveguides using electron beam lithography and plasma reactive ion etching to a depth of approximately 3,000 nm. The deep etched geometry is required to minimize the optical leakage to the high refractive index GaAs substrate. The resultant waveguide structure can be seen in Fig. 1b along with the second-order dispersion coefficients $k^{(2)}$ in (c) and (d) that dictates the device's SFWM operation bandwidth. The waveguide core is $Al_{0.25}Ga_{0.75}As$, and its aluminum concentration was chosen so that two-photon absorption (TPA) is negligible[22]; otherwise the performance of the source would be limited[23]. We have confirmed that indeed TPA is absent in our waveguides in a correlated photon pair generation experiment via a single SFWM process[24]. The waveguide widths vary from 700 nm to 1,200 nm and have a length of 4.5 mm. The waveguides are tapered at both ends to 2,000-nm-wide waveguides to facilitate in/out coupling. Propagation losses of such waveguides were measured to be from 10 to 15 dB cm$^{-1}$. In these structures the waveguide widths have a considerable effect on the dispersion of the TE mode due to the high index contrast (between air and $Al_{0.25}Ga_{0.75}As$) and high mode confinement. The zero-dispersion wavelength (ZDW) can be positioned anywhere in the entire C-band by controlling the waveguide widths[25]. This condition is desirable to achieve efficient SFWM. The ZDW is located at ~1550 nm with the width of 700 nm (see Fig. 1c, and we have shown that this waveguide delivers efficient and broadband classical four-wave mixing (FWM)[24]. On the other hand, the TM mode is only slightly affected by the waveguide widths since the major electric field aligns vertically and experience less index contrast from the core-cladding interface. Hence, it remains normally dispersive. We can estimate the bandwidth of SFWM using a classical FWM measurement[26]. Previously we reported co-polarized, continuous-wave FWM efficiencies measured from the 700-nm-wide waveguide[24]. The TE mode exhibited a bandwidth of at least 80 nm while the TM mode operated with a bandwidth of 60 nm. As the waveguide becomes wider, the dispersion property of the TE mode becomes more similar to that of the TM mode, whereas the latter is less dependent on waveguide width. This can be seen from Fig. 1c and 1d where we plot a dispersion coefficient $k^{(2)}$ as a function of wavelength for both modes in different waveguides. Therefore, it is justified to assume that all our waveguides operate with a 60-nm bandwidth for both modes in the polarization entanglement experiment.

The symmetry of the AlGaAs third-order susceptibility $\chi^{(3)}$ allows eight SFWM processes to occur inside the waveguide. The allowed processes are listed in the left column of Table 1 along with the resultant signal-idler polarization states in the right column. We denote the process using four letters with the first, second, third, and forth letters referring to polarization states of the signal, the idler, and the two pump photons. For example, the *stuv* process refers to a process with the signal and idler photons in the *s* and *t* polarizations and the two pump photons in the *u* and *v* polarizations where $s,t,u,v \in \{H,V\}$. We shall see that the processes leading to the states of $|HV\rangle$ and $|VH\rangle$ degrade the entanglement of the state we would like to generate with our scheme in equation (1).

Table 1. A list of contributing SFWM processes (left column) to corresponding signal-idler polarization states (right column). The first, second, third, and forth letters specify polarization states of the signal, idler, first pump, and second pump photons.

| SFWM processes | Signal-idler state |
| --- | --- |
| HHHH and HHVV | HH |
| VVVV and VVHH | VV |
| HVHV and HVVH | HV |
| VHHV and VHVH | VH |

**Biphoton wavefunctions and polarization-entangled states.** The properties of the entangled state of a photon pair are captured in a biphoton wavefunction (BPW). Let a waveguide span in a *z* direction from $z=-L/2$ (input facet) to $z=L/2$ (output facet), and consider a pump input state expressed as[27]

$$|\psi_{\text{in}}\rangle = \exp\left\{\sum_w \alpha_w A_w^\dagger - \text{H.c.}\right\}|0\rangle \quad (2)$$

where $A_w^\dagger = \int dk\, \phi_w(k) a_w^\dagger(k)$, $\phi_w(k)$ represents the pump spectral profile, and $w \in \{H,V\}$ denotes the mode of the pump photons (recall *H* is the TE mode and *V* represents the TM mode). The input state in equation (2) describes a superposition of coherent pump states with average numbers of photons $|\alpha_w|^2$ in the mode $w$[28]. We impose a lossless limit and focus on the situation that both the TE and TM pump modes have the same frequency, i.e. degenerate SFWM. The BPW specific to a *stuv* SFWM process (see Table 1, left column) is expressed as

$$\phi_{stuv}(k_1,k_2) = K\int dk_{s3}k_{t4}\phi_u(k_{u3})\phi_v(k_{v4})\frac{\sin(\Delta k L/2)}{\Delta k L/2}\sqrt{\omega_1\omega_2\omega_3\omega_4}\,\delta(\omega_1+\omega_2-\omega_3-\omega_4)e^{i(k_{s1}+k_{t2}+k_{u3}+k_{v4})L/2} \quad (3)$$

where *K* absorbs all constants (including an effective nonlinear strength) and $\Delta k = k_{s1}+k_{t2}-k_{u3}-k_{u4}$ is the phase mismatch of the *stuv* process. The state of the entangled signal-idler photon pair at the output of the waveguide is

$$|\psi_{gen}\rangle = \sum_{stuv} \alpha_u \alpha_v \int dk_1 dk_2 \phi_{stuv}(k_1, k_2) |s, \omega_1; t, \omega_2\rangle \quad (4)$$

After collecting the generated photon pairs, we apply bandpass filters in the state tomography measurement. Hence, the detected state can be described as

$$|\psi\rangle = \sum_{stuv} \alpha_u \alpha_v \int dk_1 dk_2 F(\omega_1, \omega_2) \phi_{stuv}(k_1, k_2) |s, \omega_1; t, \omega_2\rangle \quad (5)$$

where $F(\omega_1,\omega_2)$ represents a filter function. The generated state is a superposition of four signal-idler states $|s;t\rangle$ where each is calculated by summing over the pump polarizations $u$ and $v$ and integrating over the propagation constants of the signal and idler photons. The ratio $\alpha_v/\alpha_u \equiv r$ can be chosen from equation (5) to balance the probabilities of generating the pair in $|HH\rangle$ and $|VV\rangle$.

We plot in Figure 2 the magnitude of the four balanced BPWs $\phi_{st}$ (with summation and integration carried out) for (a) the 700-nm-wide waveguide with $r=1.33$ and (b) the 1,200-nm-wide waveguide with $r=1.14$ (see Methods for computation details and Supplementary Materials for all BPWs of the SFWM processes). The area within white rectangles corresponds to the filter function employed in the experiment and used to find the power ratio $r$. There are a couple of important features to note. The probabilities of generating photon pairs with parallel polarizations are much higher than those of generating pairs with orthogonal polarizations due to the phase matching condition. This is more obvious in the 700-nm-wide waveguide for which the orthogonally-polarized states have almost-zero probability of being generated. However, the difference in the two generation probabilities proportionally depends on the differential group index (DGI) between the TE and TM modes. The DGI is 0.10 for the 700-nm-wide waveguide and drops to 0.016 in the 1,200-nm-wide waveguide. As a result, the chance of orthogonally-polarized pairs being created increases as the waveguide widens, such as the 1,200-nm-wide waveguide. These orthogonally-polarized states then make the generated state partially factorizable and reduce the entanglement. On the other hand, a relatively larger DGI in the 700-nm-wide waveguide could lead to a significant temporal walk-off and renders the two processes partially distinguishable and downgrades the entanglement. Hence, the dispersion property manifests into two competing effects between the narrow and the wide waveguides: temporal walk-off and factorizability in the polarization degree of freedom.

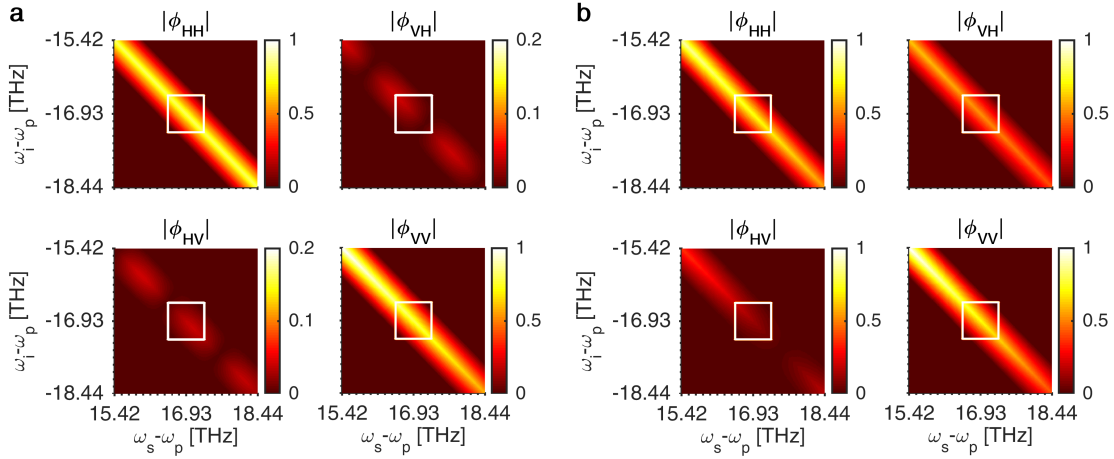

Figure 2. Magnitudes of effective, balanced biphoton wavefunctions as a result of two orthogonally pumped SFWM processes in (a) the 700-nm-wide waveguide and (b) the 1,200-nm-wide waveguide.

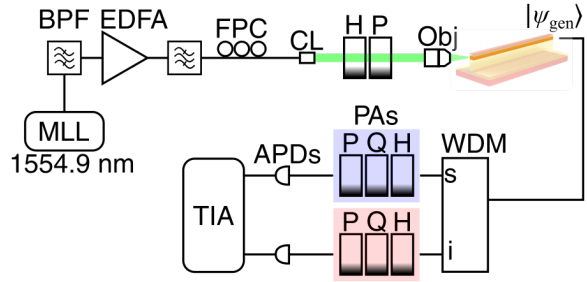

Figure 3. A schematic diagram of the experiment including the following items: a mode-locked laser (MLL), an erbium-doped fiber amplifier (EDFA), 1-nm band-pass filters (BPF), a fiber polarization controller (FPC), a fiber collimator (CL), quarter-wave plates (Q), half-wave plates (H), polarizers (P), an objective lens (Obj), a set of WDM filters, polarization analyzers (PA), avalanche photodiodes (APDs), and a time interval analyzer (TIA).

**Observation of polarization-entangled states.** The experimental setup is displayed in Fig. 3 (see Methods). We excite both the TE and TM modes with pump pulses (1,554.90 nm) prepared to be linearly polarized at ~45° to the horizontal. The photon pair is then created inside the waveguide. The output light is collected by a lensed fiber and passes through a cascade of filters (WDM in Fig. 3) that remove pump photons and separate the signal (1,533.47 nm) and the idler (1,577.03 nm) photons, each with a passband of 1 nm. Polarization controllers are calibrated and configured to unwind the fiber's polarization scrambling effect and to project the photons onto different polarization states: *H*, *V*, *D* (diagonal, -45°), *A* (anti-diagonal, 45°), *R* (right-handed circular), and *L* (left-handed circular). These polarization states are defined with respect to the lab frame. The photons are then detected using avalanche photodiodes over an integration time of 180 s. We then perform projective state tomography on the generated state with an over-complete set of signal-idler polarization states, i.e. conducting 6×6 measurements. The 36 accidental-subtracted coincidence counts are subsequently used to reconstruct a density matrix of the state in the basis $\{|HH\rangle,|VH\rangle,|HV\rangle,|VV\rangle\}$ using the maximum likelihood technique[29]. A thousand instances of Monte Carlo simulation based on the Poissonian nature of the detection are used to estimate errors of the reconstructed density matrices and their related quantities.

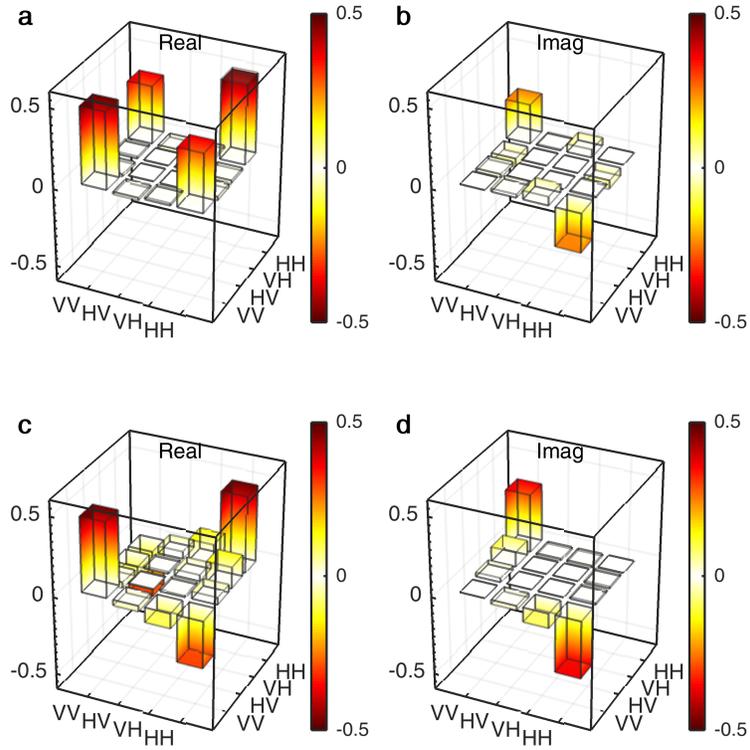

Figure 4. Reconstructed density matrices of the polarization-entangled states generated from the 700-nm-wide waveguide ((a)-(b)) and the 1,100-nm-wide waveguide ((c)-(d)).

We conducted experiments on several deeply etched AlGaAs waveguides with different widths from 700 nm to 1,200 nm. For each waveguide, the powers of the pump modes were adjusted so that the states $|HH\rangle$ and $|VV\rangle$ are detected at approximately the same rates: this is equivalent to an on-chip rate of $16.4\times10^{-3}$ pairs per pulse (see Methods). In Fig. 4, we show the reconstructed density matrices of the entangled states generated from the waveguides with widths of 700 nm ((a)-(b)) and 1100 nm ((c)-(d)). Density matrices of all other waveguides can be found in the Supplementary Material. We clearly observed four corner pillars of the density matrices, which are an expected trait of the targeted entangled state in equation (1).

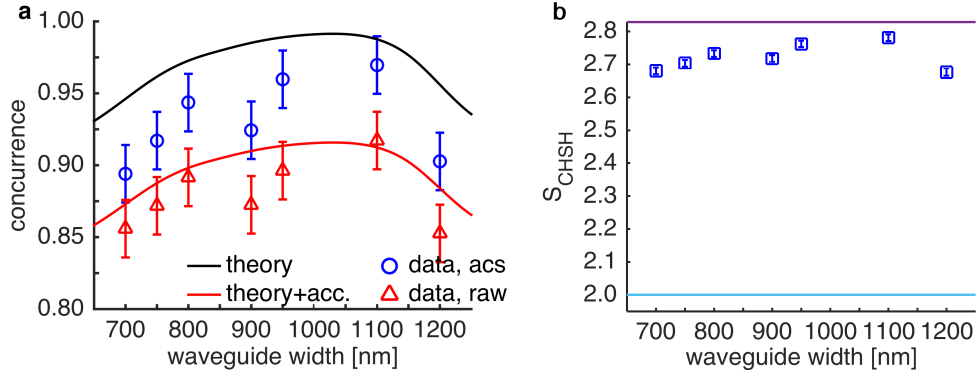

Figure 5. (a) The experimental and simulated concurrences of the generated states versus waveguide widths. The higher concurrence shown has the accidental counts subtracted while the lower concurrence shown is based on raw experimental data. (b) The expected $S$ parameter extracted from the reconstructed density matrix showing a violation of the Bell-CHSH inequality. The torquise line is the classical correlation limit whereas the magenta line marks the quantum correlation limit.

The concurrence $C$, which is a measure of the degree of entanglement, is extracted from the reconstructed density matrices and shown in Fig. 5a. The result is shown in red triangles for concurrence evaluated from the total raw coincidence counts and also in blue circles when the accidental counts are subtracted (abbreviated as acs.). The accidental counts are taken from the coincidence counts registered at the next pump pulse. The error bars represent the 95% confidence interval with a standard deviation of 0.01. Albeit being the most efficient for all-TE FWM interaction, the 700-nm-wide waveguide yielded the lowest raw concurrence at 0.85±0.01 (0.89 acs.). The measured concurrence generally increases toward wider waveguides. The maximum raw concurrence is observed from the 1,100-nm-wide waveguide with a value of $C$=0.91±0.01 (0.97 acs.), whose corresponding density matrix is shown in Fig. 4c and 4d. (All BPWs of this waveguides are plotted in Supplementary Fig. 2). The raw concurrence then drops back to 0.85±0.01 (0.90 acs.) at the waveguide width of 1,200 nm. Nonetheless, the measured results lie among the best-reported concurrences in integrated devices[12,19]. We also note that no additional setups were used here to erase any possible distinguishability.

A simulated concurrence is calculated using the model discussed previously. The solid black line in Fig. 5a represents the predicted pure state while the red curve incorporates predicted accidental counts and other experimental parameters (See Supplementary). The theory curve peaks near the width of 1,050 nm, and its shape agrees well with the experimental data. To the left of the peak, the entanglement drops as a result of the PMD that introduces distinguishability. On the other hand, to the right of the peak, the entanglement is degraded as the PMD decreases, growing contributions from the $|HV\rangle$ and $|VH\rangle$ such that the resulting state becomes more factorizable in the polarization degree of freedom. Hence, we confirm from both simulation and experiment that an optimal level of PMD is required to produce a state with high polarization entanglement. We note that accidental counts appreciably affect the resultant concurrence. At the employed pump power, the accidental counts is dominated by the chance of detecting signal and idler photons from different pairs rather than by the dark counts of the detectors.

From the accidental-subtracted reconstructed density matrices, we evaluate the violation of the Bell-CHSH inequality[30]. The $S$ parameter is defined as $|E(a,b)+E(a',b)+E(a,b')-E(a',b')|$ and it strongly depends on the two sets of measurement: $\{a,a'\}$ on one photon and $\{b,b'\}$ on the other. The correlation function $E(a,b)$ is computed from $\text{Tr}\{\rho \cdot m(a) \otimes m(b)\}$ where $m(x)$ represents a single-qubit measurement operator projecting onto the $x$ state. We numerically search for the two measurement sets that maximize the parameter $S$ for each of the Monte Carlo instances. The average of the maximized $S$ parameters is collected for each waveguide and shown in Fig. 5b with a standard deviation of 0.01. The $S$ parameter is well above 2 (red solid line) that is dictated by the classical correlation limit, and its maximum value is 2.78±0.01 from the 1100-nm-wide waveguide. It also varies linearly with the concurrence as expected.

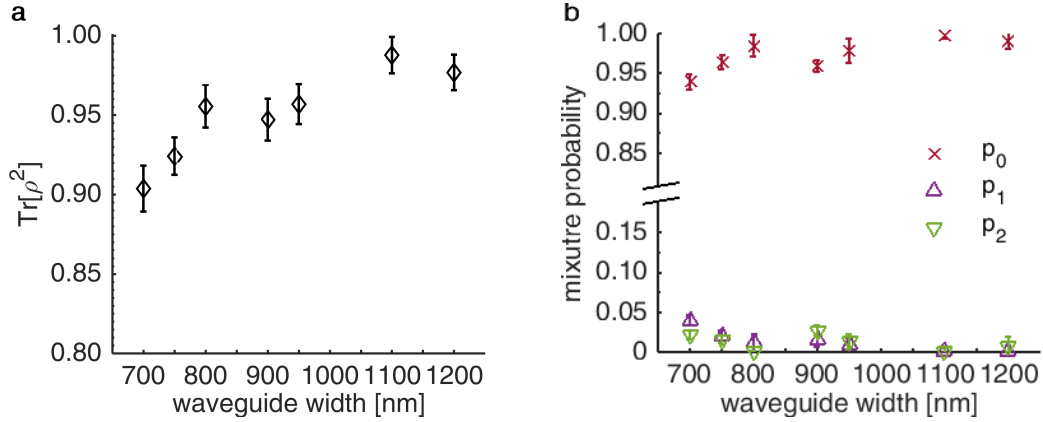

Figure 6. (a) Purity of the generated states. (b) Mixture probabilities of the impurities in the observed density matrix.

We investigate further the deviation of the measured concurrence from the theoretical line in the case when the accidental counts are subtracted. The purity of the measured state defined as $\text{Tr}[\rho^2]$, plotted in Fig. 7a, suggests that the generated state becomes more *mixed* as the waveguide becomes narrower. We attempt to decompose the measured density matrix to reveal the statistical mixture of states according to $\rho=\Sigma_i p_i \rho_i$ where $i \in \{0,1,2\}$ and $\rho_0=|\psi_{\text{gen}}\rangle\langle\psi_{\text{gen}}|$ is the predicted polarization-entangled state, $\rho_1=|HH\rangle\langle HH|$, and $\rho_2=|VV\rangle\langle VV|$. The reasoning behind the inclusion of $\rho_1$ and $\rho_2$ is the possibility of distinguishing which process creates the photon pair. An optimization search is carried out to find a set of mixture probabilities $p_i$ that maximizes the fidelity (i.e. similarity) of the search matrix and the observed density matrix with a constraint that $\Sigma_i p_i=1$. The search result is shown in Fig. 7b. The mixture probability $p_0$ from the entangled state $\rho_0$ appears as the major contribution to the overall detected state. This probability $p_0$ drops as the waveguide becomes narrower. On the other hand, the distinguishable states $\rho_1$ and $\rho_2$ have become more probable, and therefore lead to a decrease in concurrence.

We postulate that these distinguishable contributions that cause the discrepancy from the theoretical line (in Fig. 5a) originate from simulation parameters that do not perfectly represent the actual situation and from unaccounted accidental counts.

The phase offset $\theta$ between the $|HH\rangle$ and $|VV\rangle$, depends on the waveguide dispersion and the length of the device, both of which can be controlled to directly generate Bell states without additional compensation. The phase offset is extracted from the reconstructed density matrices and plotted in Fig. 8. In Fig. 9, we determine the fidelity of the generated entangled states from all the waveguides to Bell states $|\Phi^{\pm}\rangle=|HH\rangle\pm|VV\rangle$. In particular, the maximum fidelity of 0.92±0.01 (0.90 raw) to the Bell state $|\Phi^-\rangle$ is observed with the 800-nm-wide waveguide.

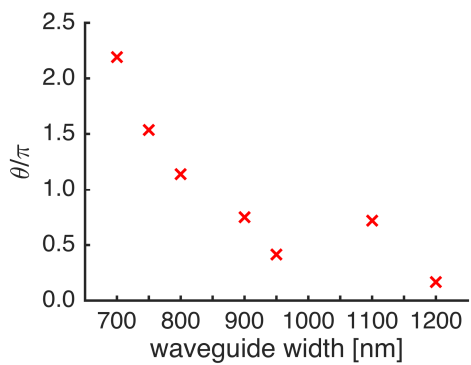

Figure 7. Relative phases of the $|VV\rangle$ with respect to the $|HH\rangle$ of the generated entangled state from different waveguides.

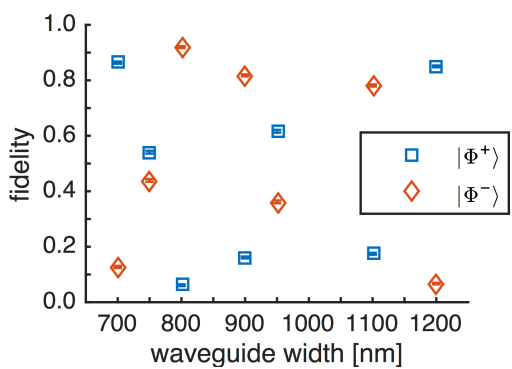

Figure 8. Fidelity of the generated states from several waveguides to the Bell states $|\Phi^{\pm}\rangle=|HH\rangle\pm|VV\rangle$, displaying a possibility of directly and tunably generating Bell states with the proposed scheme and without any post-generation compensation steps.

**Discussion**

The polarization mode dispersion between the TE and TM modes in waveguides, especially the group velocity mismatch (GVM), is usually perceived as a serious detrimental factor since it causes temporal walk-off. With sufficient propagation and GVM, the wave packets of the $|HH\rangle$ and $|VV\rangle$ states overlap less in time. As a result, it is possible to specify which process creates the photon pair by its detection time, leading to poor entanglement. Then, it might seem most desirable to have the TE and TM modes becoming identical in dispersion properties as one could achieve with square waveguides, thereby eliminating the walk-off problem. Unfortunately, in such a case, the entanglement will be completely destroyed because all of the eight SFWM processes become degenerate and similarly efficient, and the photon pair will have an equal probability of being created in each state of $\{|HH\rangle,|VH\rangle,|HV\rangle,|VV\rangle\}$. In other words, the generated state will be fully factorizable as $|\psi_{gen}\rangle = |HH\rangle+e^{i\theta s}|VH\rangle+e^{i\theta i}|HV\rangle+ e^{i(\theta s+\theta i)}|VV\rangle = (|H\rangle+e^{i\theta s}|V\rangle)_s \otimes (|H\rangle+e^{i\theta i}|V\rangle)_i$. On the other hand, in highly-asymmetric waveguide geometries, as in typical crystalline silicon waveguides, the TE and TM modes behave very differently, and thus spectral profiles of the BPWs $\phi_{HHHH}$ and $\phi_{VVVV}$ are spectrally different[17–21]. This leads to obvious distinguishability due to temporal walk-off. In this work, we have shown that an optimal level of asymmetry between the TE and TM modes is mandatory in preparing highly-polarization-entangled photon pairs using orthogonally-pumped SFWM processes. The dispersion property, waveguide length, and filtering bandwidth collaborate to simultaneously optimize the temporal walk-off and polarization state factorizability. In addition, the orthogonally pumped SFWM scheme allows one to directly generate Bell states by tuning the waveguide geometries (hence, dispersion) and waveguide lengths.

Compared to $\chi^{(2)}$-based AlGaAs devices[12–16], our sources are much simpler to fabricate due to a more relaxed phase matching requirement, and they can generate states with a high level of polarization entanglement, in fact performing at par with these conventional AlGaAs sources. The cross-sectional structure of our AlGaAs sources provides mild PMD between the fundamental TE and TM modes such that the effect of GVM is small. This is in contrast to crystalline silicon waveguides where the two modes behave very dissimilarly due to greatly asymmetric waveguide geometries. The generation of polarization-entangled photon pairs in silicon waveguides, then, can only rely on the TE polarization mode, and must require an additional polarization rotation step, such as an integrated polarization rotator[19] or a Sagnac loop[17]. The former introduces complexity in the fabrication while the latter faces with a sophisticated setup. Here, we have shown that we could circumvent these complexities by employing suitable-PMD waveguide structure. We compare the performance of our source with other demonstrations in Table 2. Our source indeed outperforms silicon-based devices and can compete with the conventional $\chi^{(2)}$-based AlGaAs sources.

Table 2. Comparison of concurrence and S parameter of our source with other sources of polarization-entangled photons.

|  | Process | Material | $C_{raw}$ | $C_{acs.}$ | $S_{acc. sub.}$ |
|---|---|---|---|---|---|
| This work | $\chi^{(3)}$, dual | AlGaAs | 0.91 ± 0.01 | 0.97 ± 0.01 | 2.78 ± 0.01[a] |
| Lv et al[21] | $\chi^{(3)}$, dual | Si | - | - | 2.59 ± 0.12[b] |
| Matsuda et al[19] | $\chi^{(3)}$, single | Si | 0.88 ± 0.02 | - | - |
| Olislager et al[18] | $\chi^{(3)}$, single | Si | - | - | 2.37 ± 0.19 |
| Kang et al[31] | $\chi^{(2)}$ | AlGaAs | 0.92 ± 0.03 | 0.98 ± 0.01 | 2.65[b] |
| Orieux et al[15] | $\chi^{(2)}$ | AlGaAs | 0.68 ± 0.01 | 0.75 ± 0.03 | 2.23 ± 0.01 |

<sup>a</sup>Estimated from reconstructed density matrices as described previously. <sup>b</sup>Estimated from reported visibility of two-photon interference visibilities.

In summary, we demonstrated an on-chip source of polarization-entangled photon pair states without extra post-generation, off-chip steps for compensating or erasing the 'which process' information. The scheme is based on orthogonal-pumping SFWM and it liberates us from complexities typically encountered in conventional $\chi^{(2)}$-based AlGaAs devices and high PMD silicon waveguides. We also showed that a moderate amount of PMD is required to produce high polarization entanglement. Our devices are deeply etched, 4.5-mm-long AlGaAs waveguides whose compositions are tailored to suppress TPA. The waveguides support two orthogonal fundamental modes with weak PMD and polarization-dependent losses, and both modes have SFWM operation bandwidths of 60 nm and could potentially accommodate about 20 channels for entanglement multiplexing[32]. We conducted full state tomography and observed raw concurrences more than 0.85±0.01 across several waveguide widths and achieved a high entanglement state with a maximum raw concurrence of 0.91±0.01 (0.97 acs.). In addition, we directly generated Bell states with a high fidelity. We note that the choice of AlGaAs materials would enable us to integrate other photonic elements to realize an integrated quantum photonic circuit. In particular, two-photon interference has been demonstrated with an integrated Mach-Zehnder interferometer implemented in GaAs[33]. Integrating superconducting single-photon detectors in the AlGaAs system has shown promise[34,35]. Together with the aforementioned integrated interferometer and detection components, we believe that our source and its insight, by greatly simplifying generation of polarization-entangled states, will significantly contribute to the realization of polarization-protocol quantum photonic circuits.

## Methods

**Simulations**. The computation of the polarization entangled state in our scheme requires the information of the waveguide dispersion, the pump spectra, and the filter function according to equations (3) and (4). The waveguide dispersion curve is determined from Lumerical MODE software. The phase matching function is computed with the full dispersion curve, i.e. no approximation is used. The pump spectra for both the excited TE and TM modes are assumed identical, justified from the experiment implementation. The spectra assume a rectangular waveform with a bandwidth of 100 GHz (~1 nm) to reflect filtering in the pump preparation step. The central wavelength of the pump is 1554.9 nm, corresponding to the central wavelength of the deployed WDM filters. The signal-idler filter function $F(\omega_1,\omega_2)$ is a 2-dimensional rectangle waveform with a bandwidth of 100 GHz (~1 nm) in both dimensions, again to reflect the filtering used in the experimental setup. The center of the function corresponds to the central wavelengths of the signal and idler filtering channels, i.e. 1,533.47 nm and 1,577.01 nm, respectively. The integral is computed with respect to an angular frequency $\omega$ instead of a propagation constant $k$ by utilizing a one-to-one relationship (with appropriate pre-factors[27]) between the two quantities due to the waveguide structure.

**Experiments**. The pump pulse comes from a 700-fs mode locked fiber laser with a pulse repetition rate of 10 MHz. The pump passes through one 1-nm-bandpass fiber-based filter with a central wavelength of 1554.90 nm to extend the temporal duration to ~6 ps. The pump pulse is then amplified with an erbium-doped fiber amplifier (EDFA) and subsequently passes two additional similar filters to remove amplified spontaneous emission (ASE) photons. The pump is later coupled out to free space with a collimator. Before the collimator, a fiber polarization controller is used to adjust the polarization of the pump in free space to be linearly polarized and aligned horizontally with respect to the lab frame. A half-wave plate (HWP) and a polarizer (P) allow further control of the polarization state of the pump. To excite both the TE and TM waveguide modes, the polarization of the pump is initially set to ~45° with respect to the horizontal. The pump is then coupled into the device via a 40× objective lens that provides a power coupling of -10 dB. At the end of the waveguide, light is collected by a tapered fiber with efficiency of -6 dB. A set of WDM filters removes pump photons with a rejection ratio of <110 dB and separates the generated signal and idler photons. In each channel, three WDM filters are connected in series with central wavelength of 1,533.47 nm for the signal and 1,577.01 nm for the idler, and effective filtering bandwidth is ~ 1 nm. Each of the separated entangled photons then enters a polarization analyzer (PA) that consists of a half-wave plate, a quarter-wave plate (QWP), and a polarizer, in sequence. The configuration of the PAs is discussed below. After exiting the PA, each of the photons is then detected with an InGaAs avalanche photodiode whose detection efficiency is configured to 20% and detection dead time is set at 15 μs. Detection clicks (both singles and double counts) from the two avalanche photodiodes are recorded with an FPGA-based time interval analyzer whose temporal resolution is 500 ps. The overall detection efficiency (starting from the chip-fiber coupling to the detector) is about -20 dB for the signal channel and -18 dB for the idler one. All fiber components in the output path are secured and stabilized to ensure minimal polarization drifting throughout a single 36-measurement tomography. We tested that the mechanical and temperature stability is maintained and the drift (i.e. change in transmitted power through the PA) is low at least within 6 hours from PA configuration step. With the correctly configured PAs, the pump-path polarizer is tuned such that the coincidence counts of two projective measurements onto the $|HH\rangle$ and $|VV\rangle$ are approximately the same. Then, a 36-measurement tomography could be conducted. This calibration process is repeated for different waveguides. The projections to the $|HH\rangle$ and $|VV\rangle$ states register total coincidence counts at a rate of ~13.3 Hz, corresponding to the on-chip photon pair generation rate of $16.4\times10^{-5}$ pairs per pulse per GHz.

**Configuring polarization analyzers**. The configurations of the PAs (i.e. orientations of their internal elements) to perform projective state tomography are found from a similar setup used in the entanglement experiment but with a slight modification. A narrow-line width, continuous-wave laser is used instead and without the EDFA. The waveguide sample is removed to isolate polarization-scrambling effects from all of the elements after the waveguide. Free-space HWP and QWP are introduced in the pump path between the polarizer and the input objective lens such that all the six polarization states ($H$, $V$, $D$, $A$, $R$, $L$) can be generated and later coupled into the tapered fiber. For each filter channel, an optical power is measured before the PA as a reference with the laser wavelength appropriately set to the channel-corresponding central wavelength. While the PA's polarizer is fixed, settings of the PA's HPW and QWP are sought such that the transmitted power through the PA is maximized for each input polarization state, i.e. achieving projection.

1. Yin, J. *et al.* Quantum teleportation and entanglement distribution over 100-kilometre free-space channels. *Nature* **488,** 185–8 (2012).
2. Wang, X.-L. *et al.* Quantum teleportation of multiple degrees of freedom of a single photon. *Nature* **518,** 516–9 (2015).
3. Pittman, T. B., Jacobs, B. C., and Franson, J. D. Probabilistic quantum logic operations using polarizing beam splitters. *Phys. Rev. A* **64,** 62311 (2001).
4. Kok, P. *et al.* Linear optical quantum computing with photonic qubits. *Rev. Mod. Phys.* **79,** 135–174 (2007).
5. Sansoni, L. *et al.* Two-particle bosonic-fermionic quantum walk via integrated photonics. *Phys. Rev. Lett.* **108,** 10502 (2012).
6. Crespi, A. *et al.* Anderson localization of entangled photons in an integrated quantum walk. *Nat. Photonics* **7,** 322–328 (2013).
7. Ladd, T. D. *et al.* Quantum computers. *Nature* **464,** 45–53 (2010).
8. Akhlaghi, M. K., Schelew, E. and Young, J. F. Waveguide integrated superconducting single-photon detectors implemented as near-perfect absorbers of coherent radiation. *Nat. Commun.* **6,** 8233 (2015).
9. Silverstone, J. W. *et al.* Qubit entanglement between ring-resonator photon-pair sources on a silicon chip. *Nat. Commun.* **6,** 7948 (2015).
10. Nambu, Y. *et al.* Generation of polarization-entangled photon pairs in a cascade of two type-I crystals pumped by femtosecond pulses. *Phys. Rev. A* **66,** 33816 (2002).
11. Grice, W. P. and Walmsley, I. A. Spectral information and distinguishability in type-II down-conversion with a broadband pump. *Phys. Rev. A* **56,** 1627–1634 (1997).
12. Horn, R. T. *et al.* Inherent polarization entanglement generated from a monolithic semiconductor chip. *Sci. Rep.* **3,** 2314 (2013).
13. Günthner, T. *et al.* Broadband indistinguishability from bright parametric downconversion in a semiconductor waveguide. *J. Opt.* **17,** 125201 (2015).
14. Autebert, C. *et al.* Integrated AlGaAs source of highly indistinguishable and energy-time entangled photons. *Optica* **3,** 143 (2016).
15. Orieux, A. *et al.* Direct Bell states generation on a III-V semiconductor chip at room temperature. *Phys. Rev. Lett.* **110,** 160502 (2013).
16. Sarrafi, P. *et al.* High-visibility two-photon interference of frequency-time entangled photons generated in a quasi-phase-matched AlGaAs waveguide. *Opt. Lett.* **39,** 5188–91 (2014).
17. Takesue, H. *et al.* Generation of polarization entangled photon pairs using silicon wire waveguide. *Opt. Express* **16,** 5721 (2008).
18. Olislager, L. *et al.* Silicon-on-insulator integrated source of polarization-entangled photons. *Opt. Lett.* **38,** 1960–2 (2013).
19. Matsuda, N. *et al.* A monolithically integrated polarization entangled photon pair source on a silicon chip. *Sci. Rep.* **2,** 817 (2012).

20. Suo, J. *et al.* Generation of hyper-entanglement on polarization and energy-time based on a silicon micro-ring cavity. *Opt. Express* **23,** 3985–95 (2015).
21. Lv, N. *et al.* 1.5 µm polarization entanglement generation based on birefringence in silicon wire waveguides. *Opt. Lett.* **38,** 2873–6 (2013).
22. Aitchison, J. S. *et al.* The nonlinear optical properties of AlGaAs at the half band gap. *IEEE J. Quantum Electron.* **33,** 341–348 (1997).
23. Husko, C. A. *et al.* Multi-photon absorption limits to heralded single photon sources. *Sci. Rep.* **3,** 3087 (2013).
24. Kultavewuti, P. *et al.* Correlated photon pair generation in AlGaAs nanowaveguides via spontaneous four-wave mixing. *Opt. Express* **24,** 3365–3376 (2016).
25. Meier, J. *et al.* Group velocity inversion in AlGaAs nanowires. *Opt. Express* **15,** 12755 (2007).
26. Fang, B. *et al.* Fast and highly resolved capture of the joint spectral density of photon pairs. *Optica* **1,** 281 (2014).
27. Helt, L. G., Liscidini, M. and Sipe, J. E. How does it scale? Comparing quantum and classical nonlinear optical processes in integrated devices. *J. Opt. Soc. Am. B* **29,** 2199 (2012).
28. Yang, Z., Liscidini, M. and Sipe, J. E. Spontaneous parametric down-conversion in waveguides: A backward Heisenberg picture approach. *Phys. Rev. A* **77,** 33808 (2008).
29. Altepeter, J. B., Jeffrey, E. R. and Kwiat, P. G. Photonic state tomography. *Adv. At. Mol. Opt. Phys.* **52,** 105–159 (2005).
30. Clauser, J. F. *et al.* Proposed experiment to test local hidden-variable theories. *Phys. Rev. Lett.* **23,** 880–884 (1969).
31. Kang, D., Kim, M., He, H. and Helmy, A. S. Two polarization-entangled sources from the same semiconductor chip. *Phys. Rev. A* **92,** 13821 (2015).
32. Lim, H. C. *et al.* Wavelength-multiplexed distribution of highly entangled photon-pairs over optical fiber. *Opt. Express* **16,** 22099 (2008).
33. Wang, J. *et al.* Gallium arsenide (GaAs) quantum photonic waveguide circuits. *Opt. Commun.* **327,** 49–55 (2014).
34. Sprengers, J. P. *et al.* Waveguide superconducting single-photon detectors for integrated quantum photonic circuits. *Appl. Phys. Lett.* **99,** 181110 (2011).
35. Reithmaier, G. *et al.* Optimisation of NbN thin films on GaAs substrates for in-situ single photon detection in structured photonic devices. *J. Appl. Phys.* **113,** 143507 (2013).

## Acknowledgements

We are thankful to the National Sciences and Engineering Research Council of Canada (NSERC) and the Edward S. Rogers Sr. Graduate Scholarship for funding.

## Author contributions

P.K. developed the entanglement concept, designed the sample, simulated performance, set and conducted experiment, and analyzed results. E.Y.Z. made the TIA unit and provided backbone codes. V.P. and M.S. fabricated the sample. X.X. helped solving major technical issues. L.Q. and J.S.A provided equipment, guidance, and data interpretation. P.K. wrote the manuscript and all other authors provided comments.

## Additional information

**Competing financial interests**: The authors declare no competing financial interests.